\documentclass[prd,twocolumn,showpacs,preprintnumbers,amsmath,amssymb,superscriptaddress,floatfix,nofootinbib]{revtex4}

\usepackage{graphicx}
\usepackage{amsmath}
\usepackage{amsfonts}
\usepackage{amssymb}
\usepackage{color}
\usepackage{multirow}
\usepackage[colorlinks, citecolor=blue,anchorcolor=red,menucolor=red, linkcolor=red,filecolor=red,runcolor=red,urlcolor=blue,frenchlinks=red]{hyperref}

\begin{document}

\title{Study on the reaction of $\gamma p \to f_1(1285) p$ in Regge-effective Lagrangian approach}

\author{Yan-Yan Wang, Li-Juan Liu, En Wang\footnote{Corresponding Author: wangen@zzu.edu.cn}, De-Min Li\footnote{Corresponding Author: lidm@zzu.edu.cn}}
\affiliation{Department of Physics, Zhengzhou University, Zhengzhou, Henan 450001, China}

\begin{abstract}

The production of the $f_1(1285)$ resonance in the reaction of $\gamma p \rightarrow f_1(1285) p$ is investigated within a Regge-effective Lagrangian approach. Besides the contributions of the $t$-channel $\rho$ and $\omega$ trajectories exchanges, we also take into account the contributions of $s/u$-channel $N(2300)$ terms, $s/u$-channel nucleon terms, and the contact term. By fitting to the CLAS data, we find
that the $s$-channel $N(2300)$ term plays an important role in this reaction. We predict the total cross section for this reaction, and find a clear bump structure around $W=2.3$~GeV, which is associated with the $N(2300)$
state. The reaction of $\gamma p \to f_1(1285) p$ could be useful to further study of the $N(2300)$ experimentally.

\end{abstract}
\date{\today}
 \maketitle

\section{Introduction}{\label{Introduction}}
It has been known that the nucleon is a bound state of three valence quarks since 1970s. Many nucleon resonances, referred as $N^*$, have been observed~\cite{Olive:2016xmw}, the properties of nucleon resonances are the important issues in hadron physics, and attract lots of attentions~\cite{Lutz:2015ejy,Aznauryan:2012ba,Klempt:2009pi}.
For the nucleon resonances with masses below 2 GeV, their properties have been widely investigated in literature. However, the current knowledge on the properties of excited nucleon  states with masses above 2 GeV is scarce. On the other hand, many missing $N^*$s, predicted by the constituent quark models are not yet found~\cite{Capstick:2000qj}.

Recently, the CLAS Collaboration has measured the $f_1(1285)$ meson for the first time in photoproduction from a proton target, and
presented the $f_1(1285)$ differential photoproduction cross section into $\eta \pi^+\pi^-$ final states from the threshold up to a center-of-mass (c.m.) energy of $W=2.8$~GeV~\cite{Dickson:2016gwc}. A cross section comparison for $\gamma p \to \eta'(958) p \to \eta \pi^+\pi^- p$ and $\gamma p \to f_1(1285) p \to \eta \pi^+\pi^- p$ at $W=2.55$~GeV in Fig.~10 of Ref.~\cite{Dickson:2016gwc} shows that the $\eta'(958)$ cross section exhibits much stronger $t$- and $u$-channel signatures in the angle dependence than dose the one of $f_1(1285)$, which is quite flat. This may imply that the $f_1(1285)$ photoproduction mechanism is not dominated alone by $t$-channel production processes.

 Before the experimental study on $\gamma p \to f_1(1285) p$~\cite{Dickson:2016gwc}, there are several theoretical works on this reaction.
 Within the Regge-model, considering the exchanges of $t$-channel $\rho$ and $\omega$ trajectories, Kochelev {\it et al.} calculated the differential cross sections of $\gamma p \to f_1(1285)p$~\cite{Kochelev:2009xz}. Comparison of the Regge-model calculations and the CLAS data shows the $t$-channel production process alone does not reproduce the CLAS measurements. Within a model motivated by Chern-Simons-term-induced interactions in holographic QCD, Domokos {\it et al.}~\cite{Domokos:2009cq} predicted the differential cross sections of $\gamma p \to f_1(1285) p$. The predictions of Ref.~\cite{Domokos:2009cq} are much smaller than the CLAS data, even in the most forward region. Based on the effective-Lagrangian approach with tree-level $\rho$ and $\omega$ exchanges in $t$-channel~\cite{Huang:2013jda}, Huang {\it et al.} presented the differential cross sections as shown in Fig. 12 of Ref.~\cite{Dickson:2016gwc}. The results of Huang {\it et al.} are also much smaller than the CLAS data.  In order to describe the CLAS data, the further model calculations are needed.

 The differences between these model predictions and the CLAS data suggest that the $s$-channel intermediate baryon resonances may play an important role in the reaction of $\gamma p \to f_1(1285)p$. That is to say, the decay of the excited $N^*$ intermediate states may be important, as pointed out by the CLAS Collaboration~\cite{Dickson:2016gwc}. The reaction of $\gamma p \to f_1(1285) p$ filters the nucleon resonances with isospin $I=1/2$, and provides a natural mode to investigate the higher excited nucleon resonance with a mass above 2.2~GeV and a sizeable coupling to the final states $f_1(1285)p$.

Among the possible nucleon resonances $N^*$ [$N(2220)$ $9/2^+$, $N(2250)$ $9/2^-$, $N(2300)$ $1/2^+$]~\footnote{According to the PDG~\cite{Olive:2016xmw}, there are three $N^*$ [$N(2220)$ $9/2^+$, $N(2250)$ $9/2^-$, $N(2300)$ $1/2^+$] in the mass range of 2.2-2.5~GeV. The differential cross sections of $\gamma p\to f_1(1285)p$ above $W=2.5$~GeV are much forward, which should be dominated by the $t$-channel mesons exchanges, as shown in Fig.~12 of Ref.~\cite{Dickson:2016gwc}, the $s$-channel nucleon resonances with masses above 2.55~GeV are not expected to give the dominant contributions.},
the $N(2300)$ can couple to the $f_1(1285)p$ in the $S$ wave, while the other two states $N(2220)$ and $N(2250)$ couple to the $f_1(1285)p$ in the $F$ and $E$ waves, respectively. It would be expected that the contributions of $E$ and $F$ waves are strongly suppressed. Thus, we will consider the state $N(2300)$ as the intermediate state in the $\gamma p\to f_1(1285)p$ reaction.

In the present work, we shall study the reaction of $\gamma p \to f_1(1285) p$ within the Regge-effective Lagrangian approach by considering the $t$-channel $\rho$ and $\omega$ trajectories exchanges, the $s/u$-channel $N(2300)$ resonance mechanisms, the $s/u$-channel nucleon terms, and contact term.

The experimental information of the two-star\footnote{According the PDG~\cite{Olive:2016xmw}, the existence evidence of the baryon states with two stars is only fair.} $N(2300)$ is very scarce\cite{Olive:2016xmw}. Until now, it was observed only in the decay of $\psi(3686)\to p \bar{N}^*(\bar{p}N^*)\to p \bar{p}\pi^0 (\bar{p}p\pi^0)$ by the BESIII Collaboration, and its mass and width are determined to be $2300^{+40}_{-30}$ MeV and $340\pm  30$ MeV, respectively~\cite{Ablikim:2012zk}. Searching for the $N(2300)$ state in other processes, for instance the photoproduction, could be useful to provide more information about the properties of $N(2300)$ state. As an isospin $1/2$ filter process, the $\gamma p\to f_1(1285)p$ is a potential mode to study the $N(2300)$ state.

This paper is organized as follows. In Sec.~II, we discuss the formalism and the main ingredients of the Regge-effective Lagrangian approach. In Sec.~III, the results and discussions are presented. Finally, a short summary is given in Sec. IV.

\section{FORMALISM AND INGREDIENTS}{\label{FORMALISM AND INGREDIENTS}}

\subsection{Feynman amplitudes}
\label{subsec:effective}

\begin{figure}[!htbp]
\includegraphics[scale=0.8]{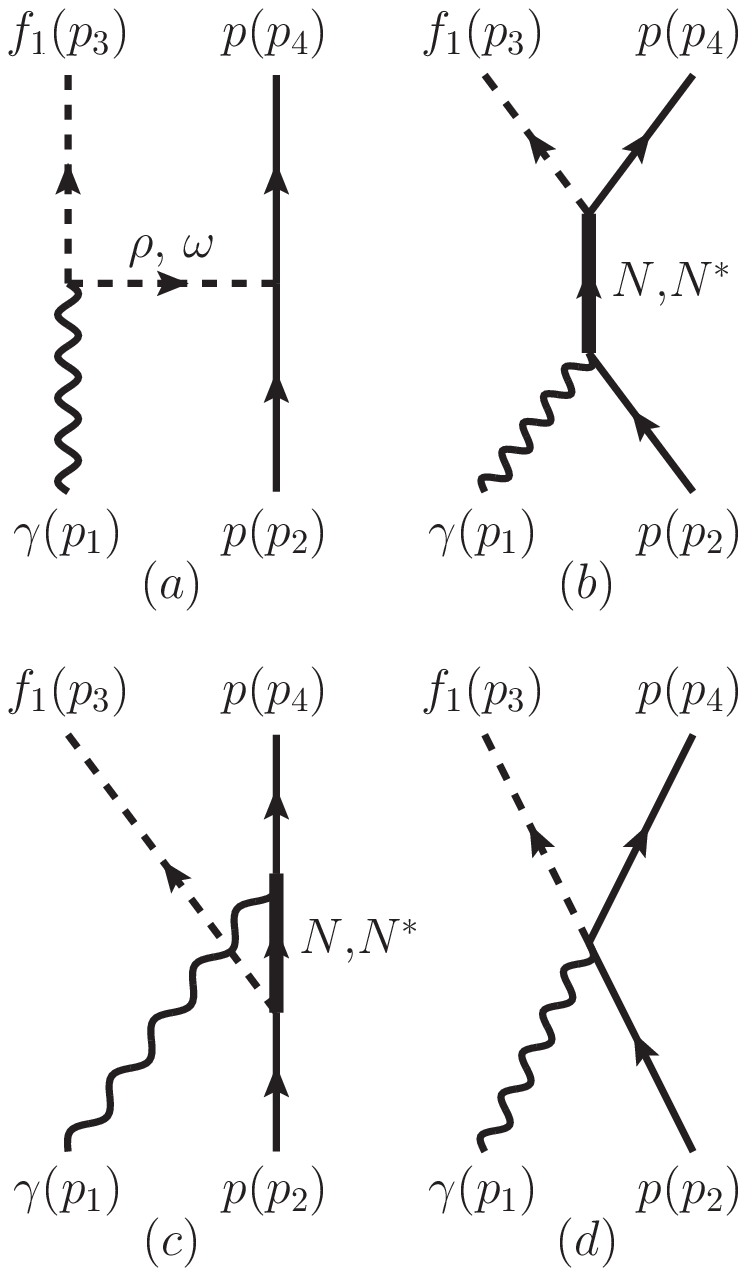}
\vspace{0.0cm} \caption{Feynman diagrams for the $\gamma p \rightarrow f_1(1285)p$ reaction: (a) the $t$-channel $\rho$ and $\omega$ exchanges; (b) the $s$-channel $N(2300)$ and proton terms; (c) the $u$-channel $N(2300)$ and proton terms; (d) contact term.}
\label{feyn}
\end{figure}

For the process $\gamma p \rightarrow f_1(1285)p$, we will take into account the basic tree level Feynman diagrams depicted in Fig.~\ref{feyn}, where the $t$-channel $\rho$ and $\omega$ exchanges, the $s$- and $u$-channel $N(2300)$ terms, the $s$- and $u$-channel nucleon terms, and contact term are considered.
The relevant effective Lagrangians of the vertices are given as~\cite{Kochelev:2009xz,Domokos:2009cq,Kochelev:1999zf,Kim:2014hha},

\begin{equation}
\mathcal{L}_{V N N} = g_{V N N} \bar{\psi}_N \gamma_{\mu} \psi_N V^{\mu} + \frac{g_V^T}{2 M_N} \bar{\psi}_N \sigma_{\mu \nu} \psi_N V^{\mu\nu}, \label{eq:lag1}
\end{equation}
\begin{equation}
\mathcal{L}_{\gamma V f_1 } = g_{\gamma V f_1 } q_V^2 \epsilon_{\mu \nu \alpha \beta} \epsilon_V^{\nu} \xi^{\beta} \epsilon_{\gamma}^{\alpha} p_1^{\mu},\label{eq:lag2}
\end{equation}
\begin{equation}
\mathcal{L}_{\gamma N N^*} = \frac{e g_{\gamma N N^*}}{2 M_{N}} \bar{\psi}_{N^*} \sigma_{\mu\nu}\partial^{\nu}A^{\mu}\psi_N + {\rm H.c.}, \label{eq:lag3}
\end{equation}
\begin{equation}
\mathcal{L}_{f_1 N N^*} = g_{f_1N N^*} \bar{\psi}_{N^*}  \gamma_5 \sigma_{\mu\nu}\partial^{\nu}f^{\mu}_1 \psi_N + {\rm H.c.},\label{eq:lag4}
\end{equation}
\begin{equation}
\mathcal{L}_{\gamma N N} = -e \bar{\psi}_{N^*}[ \gamma^{\mu}A_{\mu}-\frac{\kappa_N}{2M_N}\sigma_{\mu \nu} \partial^{\nu}A^{\mu}] \psi_N + {\rm H.c.},\label{eq:lag5}
\end{equation}

\begin{equation}
\mathcal{L}_{f_1 N N} = g_{f_1NN} \bar{\psi}_{N^*} \gamma^{\mu} \gamma_5 f_{1\mu} \psi_N + {\rm H.c.},\label{eq:lag6}
\end{equation}
where $ V^{\mu \nu} = \partial^{\mu} V^{\nu} - \partial^{\nu} V^{\mu} $, $ f_1^{\mu} $ is axial-vector meson $f_1(1285)$ field, $A^\mu$ and $V^\mu$ are electromagnetic field and the vector meson ($\rho$ or $\omega$) field, respectively.  $q_V$ is the momentum of the exchanged vector meson, and $p_i(i=1,2,3,4)$ are the four-momentum of the initial or final states, as shown in Fig.~\ref{feyn}. $\epsilon_V$, $\xi$, $\epsilon_{\gamma}$ are the polarizations of the vector meson in $t$-channel, $f_1(1285)$,  and the photon, respectively.

The numerical values of the coupling constants are taken from Ref. \cite{Sibirtsev:2004}:
\begin{equation}
g_{\omega}^T = 0,
\end{equation}
\begin{equation}
g_{\rho}^T/g_{\rho NN} = \kappa_{\rho} = 6.1.
\end{equation}

Since the hadrons are not point-like particles, we need to include the form factors to describe the off-shell effects. We adopt here the form factors used in many previous works,
\begin{equation}
\mathcal{F}_{N^*}(q^2) = \frac{\Lambda^4_{N^*}}{\Lambda^4_{N^*}+(q^2-M^2_{N^*})^2},
\end{equation}

\begin{equation}
\mathcal{F}_N(q^2) = \frac{\Lambda^4_N}{\Lambda^4_N+(q^2-M^2_N)^2},
\end{equation}

\begin{equation}
\mathcal{F}_{V N N}(q^2) = \frac{\Lambda^2_t-M_V^2}{\Lambda^2_t- q^2},
\end{equation}

\begin{equation}
\mathcal{F}_{V f_1 \gamma}(q^2) = \left(\frac{\Lambda^2_t-M_V^2}{\Lambda^2_t- q^2}\right)^2,
\end{equation}
These form factors are similar to those used in Refs.~\cite{Xie:2013mua,Xie:2010yk}, and the same cut-off $\Lambda_t$ is used for the vertices of $VNN$ and $Vf_1\gamma$.

Then, according to the Feynman rules, the scattering amplitudes for the $\gamma p \rightarrow f_1(1285)p$ reaction can be obtained straightforwardly with the above effective Lagrangians,

\begin{eqnarray}
\mathcal{M}^s_{N^*} &=& \frac{e g_{\gamma N N^*} g_{f_1 N N^*}}{2M_{N}} \mathcal{F}_{N^*}(q^2_s)
\bar{u}(p_4,s_4)\gamma_5(p_3^{\nu} - \not\!{p_3} \gamma^{\nu})\nonumber
\\&& \times G_{N^*}(q^2_s) (p_1^{\mu} - \not\!{p_1} \gamma^{\mu}) u(p_2,s_2)\nonumber
\\&& \times \varepsilon_{\mu}(p_1,s_1) \xi^*_{\nu}(p_3,s_3),
\end{eqnarray}

\begin{eqnarray}
\mathcal{M}_u^{N^*} &=& \frac{e g_{\gamma N N^*} g_{f_1 N N^*}}{2M_{N}} \mathcal{F}_{N^*}(q^2_u)
\bar{u}(p_4,s_4)(p_1^{\mu} - \not\!{p_1} \gamma^{\mu})\nonumber
\\&& \times G_{N^*}(q^2_u) \gamma_5(p_3^{\nu} - \not\!{p_3} \gamma^{\nu}) u(p_2,s_2)\nonumber
\\&& \times \varepsilon_{\mu}(p_1,s_1) \xi^*_{\nu}(p_3,s_3),
\end{eqnarray}

\begin{eqnarray}
\mathcal{M}_s^{N} &=& eg_{f_1 N N} \mathcal{F}_{N}(q^2_s)\bar{u}(p_4,s_4) \gamma^{\nu}\gamma_5 G_{N}(q^2_s)[\gamma^{\mu}-\frac{\kappa_N}{4M_N}\nonumber
\\&& \times(\gamma^{\mu}\not\!{p_1}-\not\!{p_1}\gamma^{\mu})]u(p_2,s_2) \nonumber
\\&& \times \varepsilon_{\mu}(p_1,s_1) \xi^*_{\nu}(p_3,s_3),
\end{eqnarray}

\begin{eqnarray}
\mathcal{M}_u^{N} &=& eg_{f_1NN}\mathcal{F}_{N}(q^2_u)\bar{u}(p_4,s_4) [\gamma^{\mu}-\frac{\kappa_N}{4M_N}\nonumber
\\&& \times (\gamma^{\mu}\not\!{p_1}-\not\!{p_1}\gamma^{\mu})]G_{N}(q^2_u)\gamma^{\nu}\gamma_5 \nonumber
\\&& \times u(p_2,s_2)\varepsilon_{\mu}(p_1,s_1) \xi^*_{\nu}(p_3,s_3),
\end{eqnarray}

\begin{eqnarray}
\mathcal{M}_t^V &=& - g_{V N N} g_{\gamma V f_1 } \mathcal{F}_{VNN}(q^2_t) \mathcal{F}_{Vf_1\gamma}(q^2_t) \bar{u}(p_4,s_4) \nonumber
\\&& \times \left[\gamma_{\sigma} - \frac{\kappa_{\rho}}{2 M_N}(\gamma_{\sigma}\not\!{q}_V -\not\!{q}_V \gamma_{\sigma})\right] u(p_2,s_2)\nonumber
\\&& \times G^{\sigma \nu}(q^2_t) q_V^2 \epsilon_{\mu \nu \alpha \beta} p_1^{\mu} \epsilon^{\alpha}(p_1,s_1) \xi^{*\beta}(p_3,s_3), \label{eq:amp_vector}
\end{eqnarray}
where $q_s = p_1 + p_2$, $q_u = p_2 - p_3$, and $q_t = p_1 - p_3$. $G_{N^*}$ and $G_N$ are the propagators for the $N^*$ and proton, and $G^{\sigma \nu}$ is the propagators for the $\rho$ or $\omega$ meson. We also define $g_{N^*} \equiv \sqrt{4\pi}g_{\gamma N N^*} \times g_{f_1 N N^*}$ and $g_t \equiv g_{V N N} \times g_{\gamma V f_1 }$ for convenience. With the SU(3) invariant Lagrangians and flavor symmetry, one can have $g_{\omega N N } = 3 g_{\rho N N}$. On the other hand, we have $g_{f_1 \omega \gamma } \approx \frac{e_u + e_d}{e_u - e_d} g_{f_1 \rho \gamma } = \frac{1}{3} \times g_{f_1 \rho \gamma }$~\cite{Kochelev:2009xz}, thus the $g_t$ is same for the $\rho$ and $\omega$ mesons.

The contact term is required to keep the full amplitude
gauge invariant, and can be written as

\begin{eqnarray}
\mathcal{M}_c &=& -eg_{f_1 N N} \bar{u}(p_4,s_4)\left[ \gamma^{\nu}\gamma_5 G_N(q^2_s)\not\!{p_1}\mathcal{F}_N(q^2_s) \right. \nonumber
\\&& \left. -\not\!{p_1}G_N(q^2_u)\gamma^{\nu}\gamma_5 \mathcal{F}_N(q^2_u)\right]\frac{p^{\mu}_3}{p_3 \cdot p_1}\nonumber
\\&& u(p_2,s_2)\varepsilon_{\mu}(p_1,s_1) \xi^*_{\nu}(p_3,s_3).
\end{eqnarray}

The propagator for the $N^*$ and proton term can be written as
\begin{equation}
G_{N^*}(q^2)= i\frac{\not\!{q}+ M_{N^*}}{q^2-M^2_{N^*}+iM_{N^*}\Gamma},
\end{equation}
\begin{equation}
G_{N}(q^2)= i\frac{\not\!{q}+ M_{N}}{q^2-M^2_{N}+iM_{N}\Gamma},
\end{equation}
and the one for vector meson $\rho$ or $\omega$ is
\begin{equation}
G_V^{\mu \nu}(q^2) = -i \frac{g^{\mu \nu} - q^{\mu} q^{\nu} / {M_V^2} }{q^2 - M_V^2},
\end{equation}

The total amplitude for the process $\gamma p \rightarrow f_1(1285)p$ is the coherent sum of $\mathcal{M}_s^{N^*}$, $\mathcal{M}_u^{N^*}$, $\mathcal{M}_t^{\rho}$, $\mathcal{M}_t^{\omega}$, $\mathcal{M}_s^{N}$, $\mathcal{M}_u^N$, and $\mathcal{M}_c$,
\begin{eqnarray}\label{eq:amp_eff}
\mathcal{M} &=& \mathcal{M}_s^{N^*}+\mathcal{M}_u^{N^*}+\mathcal{M}_t^{\rho}+\mathcal{M}_t^{\omega}\nonumber
\\&& +\mathcal{M}_s^{N}+\mathcal{M}_u^{N}+\mathcal{M}_c. \label{eq:amp}
\end{eqnarray}

The unpolarized differential cross section in the c.m. frame for the $\gamma p \rightarrow f_1(1285)p$ reaction is given as,
\begin{equation}
\frac{{\rm d}\sigma}{{\rm d}\Omega} = \frac{M_N^2}{16 \pi^2 s}\frac{|\vec{p}_3^{\,\rm cm}|}{|\vec{p}_1^{\, \rm cm}|} \bar{\sum}|\mathcal{M}|^2,
\end{equation}
where $s$ is the invariant mass square of the $\gamma p$ system, ${\rm d}\Omega = 2\pi {\rm d cos}\theta$, $\theta$ denotes the angle of the outgoing meson $f_1(1285)$ relative to the beam direction in the c.m. frame, while $\vec{p}^{\,\rm ~c.m.}_1$ and $\vec{p}^{\,\rm ~c.m.}_3$ are the 3-momentum of the initial photon and final $f_1(1285)$ in the c.m. frame.

\subsection{$\rho$ and $\omega$ trajectories contributions}{\label{Regge contributions}}
\label{subsec:regge}
At high energies and forward angles, Reggeon exchange mechanism plays a crucial role~\cite{Grishina:2005cy,Donnachie:1987pu}. Therefore, in modeling the reaction amplitude for the $\gamma p\to f_1(1285)p$ reaction at high energies, instead of considering the exchange of a finite
selection of individual particles, the exchange of entire Regge trajectories is taken into account, and this exchange can take place in the $t$-channel $\rho$ and $\omega$ trajectories~\cite{Wang:2014jxb}.

One can obtain the amplitude of the $\rho$ or $\omega$ trajectory exchange $\mathcal{M}_V^{\rm Regge}$ from the Feynman amplitude $\mathcal{M}_t^V$ of Eq.~(\ref{eq:amp_vector}) by replacing the usual vector meson propagator with a so-called Regge propagator~\cite{Laget:2005be},
\begin{equation}
\frac{1}{q^2 - M_V^2} \to \left(\frac{s}{s_0}\right)^{\alpha_V-1} \frac{\pi \alpha_V^{'}}{\mathrm{sin}(\pi \alpha_V)\Gamma(\alpha_V)}D_V,
\end{equation}
where the mass scale constant $s_0 = 1$ GeV, and $\alpha^{'}_V$ is the slope of the trajectory. The $\rho$ and $\omega$ trajectories are taken from Ref.~\cite{Laget:2005be},
\begin{eqnarray}
\alpha_{\omega}(t) &=& 0.44 + 0.9 t, \\
\alpha_{\rho}(t) &=& 0.55 + 0.8 t,
\end{eqnarray}
and the signature factor $D_V(t)$ is taken from Refs.~\cite{Kochelev:2009xz,Collins:1977,Guidal:1997hy}
\begin{eqnarray}
D_{\omega}(t) &=& \frac{-1 + {\rm exp}(-i \pi \alpha_{\omega})}{2}, \\
D_{\rho}(t) &=&  {\rm exp}\left(-i \pi \alpha_{\rho}\right).
\end{eqnarray}

In this work, we adopt a hybrid approach to describe the contributions of $t$-channel $\rho$ and $\omega$ exchanges in the range of laboratory photon energies explored by the CLAS data. In the hybrid approach, the amplitude $\mathcal{M}_t^V$ ($\mathcal{M}_t^\rho$ and  $\mathcal{M}_t^\omega$) in Eq.~\ref{eq:amp} is replaced by $\mathcal{M}_t^{\rm H}$~\cite{Wang:2014jxb},

\begin{eqnarray}
\mathcal{M}_t^{\rm H}=\mathcal{M}_V^{\rm Regge}\times \mathcal{R}+\mathcal{M}_t^V\times(1-\mathcal{R}), \label{eq:hybrid}
\end{eqnarray}
with
\begin{eqnarray}
\mathcal{R} &=& \mathcal{R}_s\times\mathcal{R}_t,
\end{eqnarray}
\begin{eqnarray}
\mathcal{R}_s &=&\frac{1}{1+e^{-(W-W_0)/\Delta W}},
\end{eqnarray}
\begin{eqnarray}
\mathcal{R}_t &=&\frac{1}{1+e^{(|t|-t_0)/\Delta t}},
\end{eqnarray}
where we consider $t_0$, $\Delta t$ as free parameters that will be fitted to experimental data, and $W_0 = 2.1$~GeV,$\Delta W = 0.08$~GeV from the qualitative comparison of the predictions of Ref.~\cite{Kochelev:2009xz} with the CLAS measurement~\cite{Dickson:2016gwc}, and from the findings of Ref.~\cite{Wang:2017plf}.

From the Eq.~(\ref{eq:hybrid}), we can see that for the region of high energies ($W\equiv\sqrt{s} >W_0)$ and forward angles ($|t|>t_0$), the Regge mechanism is dominant .

\section{NUMERICAL RESULTS AND DISCUSSIONS}{\label{NUMERICAL RESULTS AND DISCUSSIONS}}

 There are nine parameters in our model, (a) four relevant couplings $g_{N^*}=\sqrt{4\pi}g_{\gamma NN^*}\times g_{f_1 NN^*}$ for the $N(2300)$ term, the $g_t=g_{VNN}\times g_{\gamma V f_1}$ for the $t$-channel $\rho$ and $\omega$ trajectories exchanges, $g_{f_1NN}$ and $\kappa_N$, (b) three cut-off parameters $\Lambda_{N^*}$, $\Lambda_N$ and $\Lambda_t$, (c) $t_0$, $\Delta t$. We will obtain these parameters by fitting to the recent differential cross sections data from the CLAS experiment. Since the CLAS Collaboration presents the differential cross sections for $\gamma p\to f_1(1285)p \to \eta \pi^+ \pi^- p$, our results for the total cross section and differential cross sections have been scaled by the PDG branching fraction $\Gamma[f_1(1285)\to \eta \pi^+\pi^-]$ in the fit: $0.52\times(2/3)$, which was used by CLAS Collaboration~\cite{Dickson:2016gwc}. In our fit, $M_N=0.938$~Gev, $M_{N^*}=2.30$~GeV, $\Gamma_{N^*}=0.34$~GeV, $M_\rho=0.775$~GeV, and $M_\omega=0.783$~GeV~\cite{Olive:2016xmw}.

\begin{figure}[!htbp]
\includegraphics[scale=0.8]{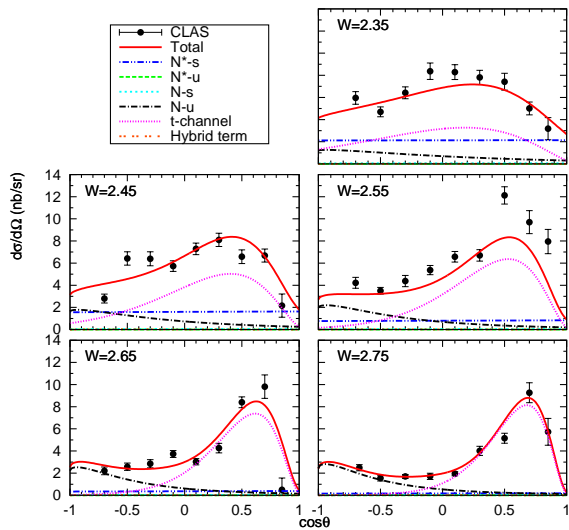}
\vspace{0.0cm} \caption{Differential cross sections of the $\gamma p \rightarrow f_1(1285) p$ reaction as a function of $\cos\theta$. The black dots are the experimental data with statistical errors~\cite{Dickson:2016gwc}. The blue dash-dot-dotted and green long dashed lines represent the contributions of the $s$-channel and $u$-channel $N(2300)$,
the black dash-dotted and cyan short-dashed lines are the contributions of the $s$-channel and $u$-channel proton, magenta dotted line describes the contribution of $t$-channel Reggeon exchanges, and orange dash-dashed line depicts the contribution of Hybrid term. The red solid line stands for the total contributions.}
\label{dcs+Nstar}
\end{figure}

There are a total of 45 CLAS experimental data, and the statistical and systematic uncertainties are taken into account. Considering all the contributions depicted in Fig.~\ref{feyn}, we perform a fit to the CLAS data, and the corresponding results  are shown in Table~\ref{fits} (Fit A).
 With these parameters in Fit A, the calculated differential cross sections from $W=2.35$ to $W=2.75$ GeV as well as the 45 available data are depicted in Fig.~\ref{dcs+Nstar}, where the blue dash-dot-dotted and magenta dotted lines correspond to the contributions of the $s$-channel $N(2300)$ and $t$-channel exchanges, respectively, the black dash-dotted is the contribution of the $u$-channel proton, and other contributions are very small, and can be neglected. The red solid lines stand for the total contributions. Only the statistical errors are shown in Fig.~\ref{dcs+Nstar}.

\begin{table}
\begin{center}
\caption{ \label{fits} The fitted parameters in this work.}
\footnotesize
\begin{tabular}{lccc}
\hline\hline
                                  & Fit A                     & Fit B             \\\hline
  $g_{N^*}$~(GeV$^{-1}$)                     & -0.052 $\pm$ 0.006            & $-$                    \\
  $g_t$~(GeV$^{-2})$           & 0.335 $\pm$ 0.068             & -0.443 $\pm$ 0.117        \\
  $g_{f_1NN}$           & 0.347 $\pm$ 0.258             & 0.110 $\pm$ 0.016     \\
  $\kappa_N$                   & -4.034 $\pm$ 2.459            & 9.999 $\pm$ 9.801    \\
  $\Lambda_{N^*}$~(GeV)        & 1.354 $\pm$ 0.269             & $-$                   \\
  $\Lambda_{N}$~(GeV)          & 1.285 $\pm$ 0.216             & 1.540 $\pm$ 0.157     \\
  $\Lambda_t$~(GeV)            & 0.582 $\pm$ 0.219             & 0.610 $\pm$ 0.187    \\
  $t_0$~(GeV$^2)$              & 2.153  $\pm$ 0.445            & 1.612 $\pm$ 0.507    \\
  $\Delta t$~(GeV$^2)$         & 0.736  $\pm$ 0.433            & 0.850 $\pm$ 0.344  \\ \hline
   $\chi^2$/dof                & 1.05                          &1.89 \\
\hline\hline
\end{tabular}
\end{center}
\end{table}

From Fig.~\ref{dcs+Nstar}, we can see that our model gives an overall reasonable description of the data in the range of $W=2.35\sim 2.75$~GeV. The $N(2300)$ provides a flat contribution for the differential cross sections, since the $N(2300)$ couples to the final states $f_1(1285)p$ in the $S$-wave. Near the threshold, the $s$-channel $N(2300)$ gives a large contribution. At higher energies, the contributions of the $t$-channel $\rho$ and $\omega$ trajectories exchanges are responsible for the shapes of the differential cross sections. The contribution of $u$-channel nucleon term is dominant at the backward angles, especially in the region of high energies. Other contributions are very small and can be neglected.

\begin{figure}[!htbp]
\includegraphics[scale=0.8]{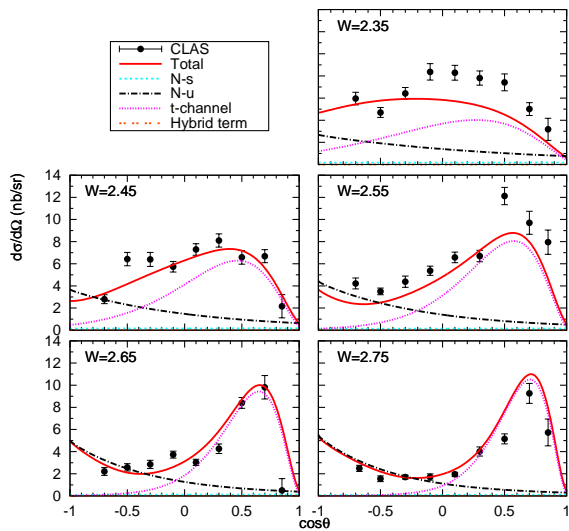}
\vspace{0.0cm} \caption{Differential cross sections of the $\gamma p \rightarrow f_1(1285) p$ reaction as a function of $\cos\theta$, where $N(2300)$ effects are not considered. The explanation is the same as that of Fig.~\ref{dcs+Nstar}.}
\label{dcs-Nstar}
\end{figure}

In order to study the role of the $N(2300)$ resonance in the $\gamma p\to f_1(1285)p$ reaction, we also perform a fit by excluding the $s/u$-channel $N(2300)$ terms and remaining the other terms.
 The corresponding results are listed in Table~\ref{fits}(Fit B), and the differential cross sections are shown in Fig.~\ref{dcs-Nstar}. From the Table~\ref{fits}, one can see that the
$\chi^2$/dof=1.89 in Fit B is larger than $\chi^2$/dof=1.05 in Fit A, which shows that the model including the $N(2300)$ contributions can better describe the CLAS data.

\begin{figure}[!htbp]
\includegraphics[scale=0.8]{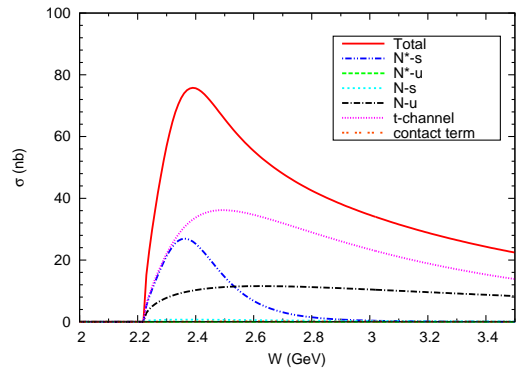}
\vspace{0.0cm} \caption{Total cross section of the $\gamma p \rightarrow f_1(1285) p$ reaction versus the invariant mass $W=\sqrt{s}$ of $\gamma p$ system, by including all the contribution depicted in Fig.~\ref{feyn}. The explanation is the same as that of Fig.~\ref{dcs+Nstar}. }
\label{tcs+Nstar}
\end{figure}

 \begin{figure}[!htbp]
\includegraphics[scale=0.8]{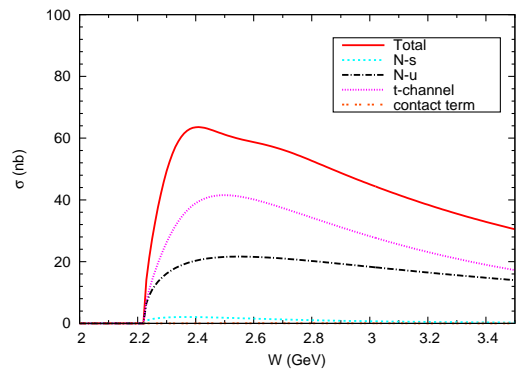}
\vspace{0.0cm} \caption{Total cross section of the $\gamma p \rightarrow f_1(1285) p$ reaction versus the invariant mass $W=\sqrt{s}$ of $\gamma p$ system, by excluding the contributions of the $s/u$-channel N(2300) terms. The explanation is the same as that of Fig.~\ref{dcs+Nstar}. }
\label{tcs-Nstar}
\end{figure}

Finally, we present the total cross section of the $\gamma p \to f_1(1285)p$ reaction with and without $N(2300)$ terms, respectively in Fig.~\ref{tcs+Nstar} and Fig.~\ref{tcs-Nstar}. From Fig.~\ref{tcs+Nstar}, it can be seen that the $s$-channel $N(2300)$ term gives a clear peak structure around $W=2.3$~GeV with the magnitude of order $30$~nb, but there is no such structure for the case of excluding the $N(2300)$ contributions, which can be used to test our model.

\section{Summary}
In this work, we have performed the study of $\gamma p \rightarrow f_1(1285) p$ reaction within the Regge-effective Lagrangian approach.
Besides the contributions from the $t$-channel $\rho$ and $\omega$ trajectories exchanges, we also consider the $s/u$-channel  $N(2300)$ terms, the $s/u$ channel of nucleon terms, and the contact term.

We extract the information about the intermediate states by fitting to the CLAS data. We find that the model including the $N(2300)$ contributions can better describe the CLAS data.



Our results indicate that
the contributions of the $u$-channel $N(2300)$ term, the $s$-channel nucleon term and contact term, are very small and can be neglected. However, the contribution of $u$-channel nucleon term is dominant at the backward angles, especially in the region of high energies.
  With the preferred parameters (Fit A), we predict the total cross section. There is a clear bump structure around $W=2.3$~GeV, which is associated with the $N(2300)$ state.
Thus, the reaction of $\gamma p \to f_1(1285) p$ could be useful to further study of the $N(2300)$ experimentally.

It should be noted that after we submitted this work to arXiv, Wang and He also discussed the $\gamma p \to f_1(1285)p$ reaction within a similar way~\cite{Wang:2017plf}, where the $t$-channel $\rho$ and $\omega$ trajectories exchanges, the $s$- and $u$-channel nucleon term are considered, and the $s$-channel nucleon resonances are not included. They suggested that the $s$-channel nucleon resonances is not very large. Our calculations show that the $s$-channel N(2300) term plays an important role in the $\gamma p\to f_1(1285) p$ reaction, and
a clear bump structure in the total cross section around $\sqrt{s}=2.3$~GeV is predicted. The current information about this reaction is not enough to distinguish our model and the one of Ref.~\cite{Wang:2017plf}.
To shed light on the relevant mechanisms of the $\gamma p\to f_1(1285)p$ reaction,
the further measurement of the total cross sections is called for.

\section*{Acknowledgements}
We would like to thank Dr. Ju-Jun Xie and Dr. Qi-Fang L\"{u} for valuable discussions. This work is partly supported by the National Natural Science Foundation of China under Grant No. 11505158, the China Postdoctoral Science Foundation under Grant No. 2015M582197, the Postdoctoral Research Sponsorship in Henan Province under Grant No. 2015023, and the Startup Research Fund of Zhengzhou University under Grants No. 1511317001 and No. 1511317002.

\end{document}